\documentclass[sigconf]{acmart}

\usepackage{algorithm}
\usepackage{algpseudocode}
\usepackage{graphicx}
\usepackage{subcaption}
\usepackage{booktabs,tabularx}
\usepackage{paralist}

\AtBeginDocument{%
  }

\setcopyright{none}
\renewcommand\footnotetextcopyrightpermission[1]{}

\copyrightyear{2026}
\acmYear{2026}
\setcopyright{cc}
\setcctype{by}
\acmConference[UMAP '26]{34th ACM Conference on User Modeling, Adaptation and Personalization}{June 08--11, 2026}{Gothenburg, Sweden}
\acmBooktitle{34th ACM Conference on User Modeling, Adaptation and Personalization (UMAP '26), June 08--11, 2026, Gothenburg, Sweden}
\acmDOI{10.1145/3774935.3806784}
\acmISBN{979-8-4007-2311-7/2026/06}
\begin{document}

\title{Vibrotactile Preference Learning: Uncertainty-Aware\\Preference Learning for Personalized Vibration Feedback}

\author{Rongtao Zhang}
\authornote{Both authors contributed equally to this research.}
\affiliation{%
  \department{Department of Computer Science} 
  \institution{University of Southern California}      
  \city{Los Angeles}
  \state{California}
  \country{USA}
}
\email{rongtaoz@usc.edu}

\author{Xin Zhu}
\authornotemark[1]
\affiliation{%
  \department{Department of Computer Science} 
  \institution{University of Southern California}      
  \city{Los Angeles}
  \state{California}
  \country{USA}
}
\email{xinzhu@usc.edu}

\author{Masoume Pourebadi Khotbehsara}
\affiliation{%
  \department{Department of Computer Science} 
  \institution{University of Southern California}      
  \city{Los Angeles}
  \state{California}
  \country{USA}
}
\email{pourebad@usc.edu}

\author{Warren Dao}
\affiliation{%
  \department{Department of Computer Science} 
  \institution{University of Southern California}      
  \city{Los Angeles}
  \state{California}
  \country{USA}
}
\email{wdao@usc.edu}

\author{Erdem Bıyık}
\affiliation{%
  \department{Department of Computer Science} 
  \institution{University of Southern California}      
  \city{Los Angeles}
  \state{California}
  \country{USA}
}
\email{biyik@usc.edu}

\author{Heather Culbertson}
\authornote{Corresponding author.}
\affiliation{%
  \department{Department of Computer Science} 
  \institution{University of Southern California}      
  \city{Los Angeles}
  \state{California}
  \country{USA}
}
\email{hculbert@usc.edu}

\renewcommand{\shortauthors}{Zhang et al.}
\begin{abstract}
Individual differences in vibrotactile perception underscore the growing importance of personalization as haptic feedback becomes more prevalent in interactive systems. We propose Vibrotactile Preference Learning (VPL), a system that captures user-specific preference spaces over vibrotactile parameters via Gaussian-process-based uncertainty-aware preference learning. VPL uses an expected information gain-based acquisition strategy to guide query selection over 40 rounds of pairwise comparisons of overall user preference, augmented with user-reported uncertainty, enabling efficient exploration of the parameter space. We evaluate VPL in a user study (N = 13) using the vibrotactile feedback from a Microsoft Xbox controller, showing that it efficiently learns individualized preferences while maintaining comfortable, low-workload user interactions. These results highlight the potential of VPL for scalable personalization of vibrotactile experiences.
\end{abstract}

\ccsdesc[500]{Information systems~Personalization}
\ccsdesc[500]{Human-centered computing~User models}
\ccsdesc[500]{Human-centered computing~User studies}
\ccsdesc[300]{Human-centered computing~Haptic devices}
\ccsdesc[100]{Computing methodologies~Active learning settings}

\keywords{vibrotactile feedback, preference learning, personalization, active learning, gaussian processes, user modeling}

\begin{teaserfigure}
  \includegraphics[width=\textwidth, trim=0cm 1.1cm 0cm 1cm, clip]{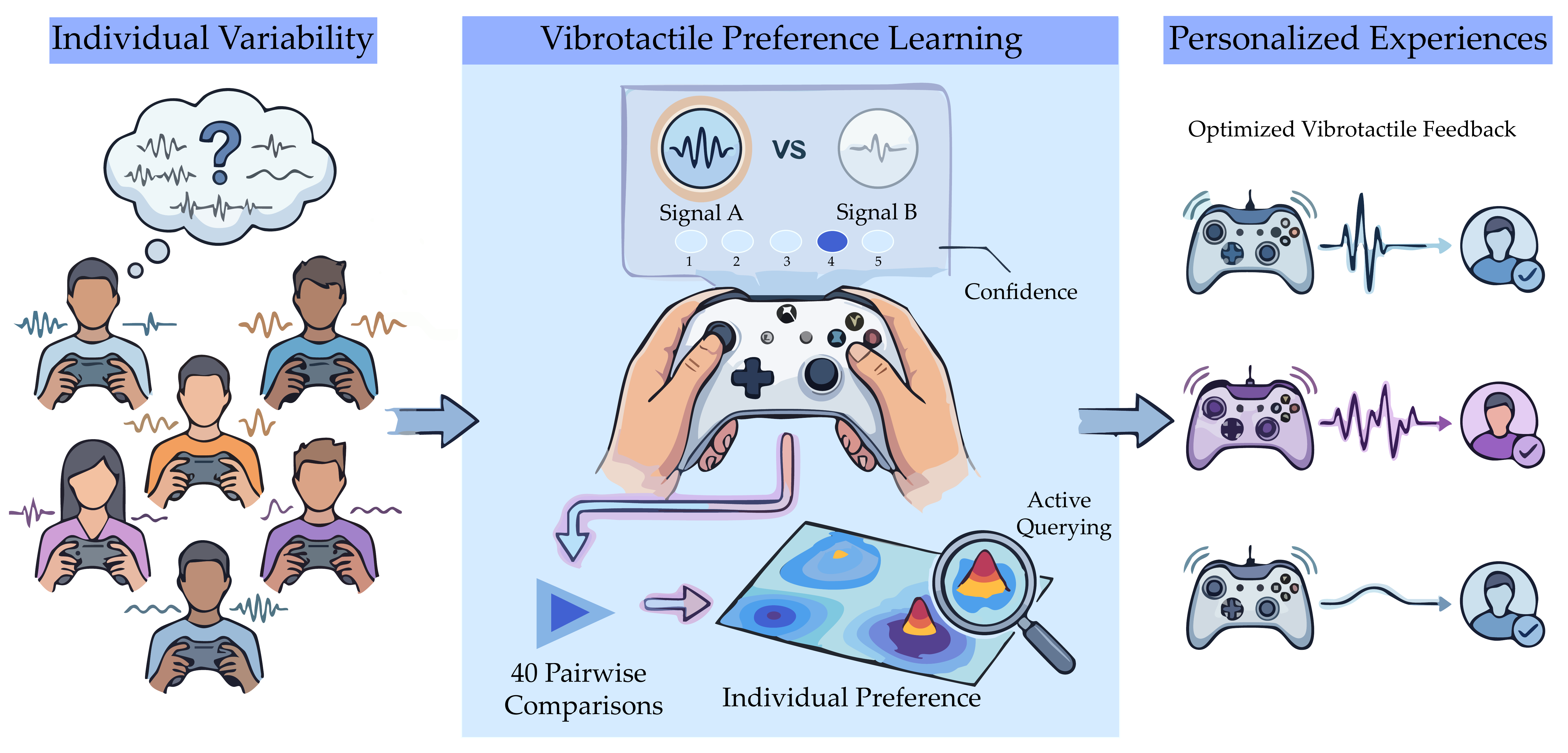}
  \caption{Overview of the Vibrotactile Preference Learning System}
  \label{fig:teaser}
\end{teaserfigure}

\maketitle

\section{Introduction}

Haptic feedback is a key channel for conveying information and enhancing immersion in virtual and remote environments~\cite{pacchierotti2017wearable,banerjee2025virtual}. Advances in haptic technology have enabled precise feedback that supports perception and manipulation of digital objects. Across mobile, wearable, and VR systems, effective haptics improve realism, reaction times, and emotional engagement~\cite{hoggan2008tactile,pacchierotti2017wearable,banerjee2025virtual}.

However, a fundamental challenge in haptic design is tactile subjectivity. It relies heavily on individual physiology (e.g., mechanoreceptor density, skin thickness) and psychology (e.g., preferences for subtle versus intense sensations). A signal designed as a ``gentle notification'' may be imperceptible to elderly users yet uncomfortable for younger ones. Consequently, predefined haptic libraries often fail to deliver the intended user experience.

To accommodate such individual differences, personalization is essential. Traditional approaches often adopt ``Method of Adjustment'' interfaces~\cite{gescheider1997psychophysics,kingdom2016psychophysics}, where users manually tune sliders for parameters such as frequency and amplitude. While flexible, this approach assumes that users understand these parameters and can effectively navigate a high-dimensional space, imposing extra cognitive load and time burdens on non-expert users. Alternatively, conventional personalization methods may ask users to rate a series of signals (e.g., via a 1--7 Likert scale~\cite{likert1932technique}). However, such methods exhibit fundamental psychometric limitations during prolonged sessions:
\begin{compactitem}
    \item \textbf{Score Drift:} Users lack a stable internal reference scale~\cite{foley1990context,kiritchenko2017bws}. A stimulus rated as ``5'' early in a session may be rated as ``3'' later as the user experiences a broader range of sensations, introducing noise into the dataset.
    \item \textbf{User Fatigue:} Providing absolute quantitative judgments for complex sensory inputs is cognitively demanding. Fatigue leads to data quality degradation and inconsistent responses~\cite{kiritchenko2017bws}.
\end{compactitem}

Humans are effective comparators, but poor absolute meters~\cite{brochu2007activepref}. While replacing absolute ratings with pairwise comparisons (A/B testing) addresses the issues of drift and fatigue, binary choice data is significantly sparser than rating scores. Furthermore, selecting which two signals to compare to maximize the informational value of each response poses a challenge in balancing personalization needs with the minimization of user effort.

Inspired by recent advances in preference learning, we seek to overcome these limitations by proposing an interactive \textbf{confidence-aware Vibrotactile Preference Learning (VPL)} system. \autoref{fig:teaser} illustrates the framework of the proposed system.

This work makes the following contributions:
\begin{compactenum}
    \item \textbf{A vibrotactile personalization preference model:}
    We use pairwise comparisons and incorporate user-reported confidence to build a Gaussian-process (GP) preference model that captures the utility of vibrotactile signals. We select the next A/B comparison by maximizing the expected information gain over vibrotactile signal parameters, enabling data-efficient learning of the global preference structure under a limited interaction budget.

    \item \textbf{Empirical validation via a user study:}
    In a user study with an Xbox controller ($N=13$), we show that VPL achieves high prediction accuracy and produces user-satisfying recommendations within 40 comparisons, aligns closely with manually tuned optima, and maintains low subjective workload.

    \item \textbf{Design insights for uncertainty-aware haptics:}
    We analyze sampling behaviors (e.g., exploration vs.\ refinement) and discuss practical implications for deploying preference-based personalization in vibrotactile settings with perceptual ambiguity and substantial individual variability.
\end{compactenum}

To comprehensively evaluate the effectiveness of the system, we investigate the following three Research Questions (RQs):
\begin{compactenum}
    \item Can the proposed confidence-aware preference learning system recover an individual’s latent vibrotactile preference structure with high predictive accuracy within a bounded interaction budget (40 comparisons)?
    \item Do users perceive the system’s recommended vibrotactile signal as matching their subjective preferences and expectations?
    \item Is the workload and time cost associated with the 40-round personalization process acceptable for short-session personalization in practice?
\end{compactenum}

We evaluated our Vibrotactile Preference Learning (VPL) system in a user study ($N=13$) using vibrotactile feedback from a Microsoft Xbox controller. Additionally, we developed a control system adapted to this controller to validate the learned preferences. The results indicate that our method achieves excellent data efficiency, converging to a stable preference model within 40 rounds. In subjective evaluations, users reported high satisfaction with the recommended signals. Furthermore, NASA-TLX results indicate that the cognitive workload remains low, with users generally finding the interaction intuitive and easy to use.
Our code is available as open-source\footnote{https://github.com/HaRVI-Lab/vibrotactile-preference-learning}.

\section{Related Work}

Haptic systems are increasingly expected to move beyond generic feedback toward \emph{user-specific} experiences, as tactile perception and comfort vary substantially across individuals, contexts, and body locations. We review prior work along three threads most relevant to our contribution: (1) personalized haptic design, (2) preference learning with uncertainty quantification, and (3) preference-driven learning in interactive systems.

\subsection{Personalized Haptic Design}
A growing body of HCI work explores personalization to improve the usability, comfort, and effectiveness of haptic systems. One line of research focuses on \emph{applied, user-facing systems}, such as \textit{Calm Commute}, which uses personalized haptic breathing guidance for stress management in drivers~\cite{balters2020calm}, and a clippable pneumatic-haptic device by Choi \emph{et al.} that allows users to tailor actuation patterns to individual comfort and body sites~\cite{choi2022design}.

Beyond applications, researchers have proposed higher-level interaction abstractions for reasoning about vibrotactile effects. Examples include Seifi \emph{et al.}’s ``affective handles'' for tuning vibrations by affective intent~\cite{seifi2018toward} and \textit{TacTalk}, which enables conversational customization of haptics via natural language~\cite{mishra2025tactalk}. Collectively, these works underscore both the importance of personalization and the need for interaction paradigms that reduce user burden while supporting exploration of complex haptic design spaces.

\subsection{Preference Learning and Uncertainty Quantification}
Preference learning replaces absolute ratings with pairwise comparisons, a paradigm rooted in psychometrics and psychophysics~\cite{thurstone1927law,bradley1952paired}. In machine learning, probabilistic models such as GP-based preference learning infer latent utilities from sparse comparisons while quantifying uncertainty~\cite{chu2005gppref}. Active preference learning improves data efficiency by selecting informative queries~\cite{brochu2007activepref,brochu2010bo}, with extensions to interactive optimization and reward learning that emphasize query efficiency under limited human feedback~\cite{biyik2024active}.

Crucially, human judgments are noisy and context-dependent. Recent work has begun to explicitly model multiple sources of uncertainty in preference learning, including human response variability~\cite{peng2025uncertainty}. Motivated by this, we incorporate user-reported confidence into GP-based preference learning to modulate trust in each comparison, while using information-gain-driven query selection under a limited interaction budget.

\subsection{Preference-Driven Learning in Interactive Systems}
Beyond estimating preferences, prior work uses preferences to \emph{drive} learning, generation, and search. In haptics, Lu \emph{et al.} demonstrated preference-driven texture modeling through interactive generation and search, showing how iterative comparisons guide systems toward user-aligned tactile outcomes~\cite{lu2022texture}. Similar interactive loops appear in preference-based reinforcement learning and reward learning, where human comparisons supervise optimization~\cite{wirth2017survey,biyik2024active}.

More broadly, preference-driven training underpins reinforcement learning from human feedback (RLHF). Foundational work learned reward models from human comparisons~\cite{christiano2017deep}, with later applications to text generation, including summarization~\cite{stiennon2020summarize} and instruction-following language models~\cite{ouyang2022instructgpt}. Across modalities, these approaches highlight human preferences as flexible supervision when explicit metrics are unavailable. Our work extends this paradigm to vibrotactile feedback, emphasizing efficient, uncertainty-aware preference acquisition for personalization.

\section{System \& Method}
\label{sec:method}

\begin{figure*}[t]
  \centering
  \includegraphics[width=\textwidth, trim=0cm 1cm 0cm 0.5cm, clip]{fig/fig2system.pdf}
  \caption{VPL Pipeline: Users make pairwise vibrotactile judgments and report confidence; a confidence-aware GP preference model updates, selects informative queries via information gain, and outputs an uncertainty-calibrated recommendation.}
  \label{fig:system_overview}
\end{figure*}

\subsection{Vibrotactile Signal Parameterization}
\label{subsec:parameterization}

We generate vibrotactile signals using a standard Microsoft Xbox controller equipped with two asymmetric actuators: a low-frequency (left) motor and a high-frequency (right) motor. To enable expressive yet manageable personalization, we parameterize the signal space using four dimensions:
\begin{equation}
\label{eq:param_ranges}
\phi \triangleq [I, T, R, G]^\top \in \Phi \subset \mathbb{R}^4,
\end{equation}
where the parameters and their hardware-constrained ranges are defined as:
\begin{compactitem}
    \item \textbf{Intensity ($I \in [0.20, 1.00]$):} The overall vibration magnitude.
    \item \textbf{Motor Balance ($T \in [0.00, 1.00]$):} The distribution of intensity between the two motors, controlling the tactile ``texture.'' The left ($M_L$) and right ($M_R$) motor strengths are computed as:
    \begin{equation}
    M_L = I \cdot T, \qquad M_R = I \cdot (1 - T).
    \end{equation}
    \item \textbf{Rhythm ($R \in [0.60, 4.00]$ Hz):} The frequency of vibration pulses. The cycle period is given by $P = 1/R$.
    \item \textbf{Grain ($G \in [0.10, 0.70]$):} The duty cycle of each pulse, affecting perceived sharpness.
\end{compactitem}

To ensure signal stability on the hardware, we enforce a minimum inter-pulse gap of 45 ms and a minimum pulse attack of 20 ms. The effective pulse duration $d$ is therefore constrained by:
\begin{equation}
d = \min\big(P \cdot G,\ \max(20\text{ms},\ P - 45\text{ms})\big).
\end{equation}
Each signal is rendered as a sequence of these pulses for a duration of 3 seconds.

\subsection{Problem Formulation}
\label{subsec:problem}

We formulate vibrotactile personalization as an interactive preference learning problem over the signal space defined above. To facilitate optimization, we linearly normalize the physical parameter space $\Phi$ to a unit hypercube $\mathcal{X} = [0,1]^4$. For the remainder of this paper, we denote a normalized query point simply as $x \in \mathcal{X}$.

We assume a latent utility function $f: \mathcal{X} \to \mathbb{R}$ governs the user's preference. The personalization process proceeds in rounds. At round $t$, the system presents a pair of signals $(x_t^A, x_t^B)$. The user provides:
\begin{compactenum}
    \item A binary preference $y_t \in \{+1, -1\}$, where $+1$ indicates $x_t^A \succ x_t^B$.
    \item A confidence score $c_t \in \{1, \dots, 5\}$ (Likert scale: $1=$ Very Unsure, $5=$ Very Sure).
\end{compactenum}
The collected dataset at time $t$ is denoted as $\mathcal{D}_t = \{(x_i^A, x_i^B, y_i, c_i)\}_{i=1}^t$. 

Our goal is to learn a user-specific latent utility function $f$ from a small number of
pairwise comparisons:
\begin{equation}
\hat f \;=\; \arg\max_{f}\; p(\mathcal{D}_T \mid f),
\end{equation}
and use the learned model to guide query selection and make recommendations.

\subsection{Uncertainty-Aware Preference Modeling}
\label{subsec:gp_pref}

\paragraph{GP prior.}
We place a Gaussian Process prior on the latent utility function
$f:\mathcal{X}\to\mathbb{R}$ over normalized parameters $x\in[0,1]^4$:
\begin{equation}
\label{eq:gp_prior}
f \sim \mathcal{GP}(0, k(\cdot,\cdot)),
\end{equation}
with an isotropic RBF kernel
\begin{equation}
\label{eq:rbf}
k(x, x') = s^2 \exp\!\left(-\frac{\|x - x'\|_2^2}{2\ell^2}\right).
\end{equation}

\paragraph{Confidence-aware probit likelihood.}
At round $i$, the user compares $(x_i^A,x_i^B)$ and returns a preference
$y_i\in\{+1,-1\}$ (where $+1$ denotes $x_i^A \succ x_i^B$) and a confidence score
$c_i\in\{1,\dots,5\}$.
Let $\Delta_i \triangleq f(x_i^A)-f(x_i^B)$.
We map confidence to a human-uncertainty noise scale
$\sigma_i \triangleq u(c_i)$ with $u(1)>\cdots>u(5)>0$,
so lower confidence corresponds to noisier comparisons.
We additionally include a small baseline jitter $\lambda>0$ to account for residual
variability not captured by confidence (e.g., actuator/control noise).
The probit likelihood is
\begin{equation}
\label{eq:probit_like}
p(y_i \mid f, c_i)
=
\Phi\!\left(
\frac{y_i \Delta_i}{\sqrt{2\lambda^2+\sigma_i^2}}
\right),
\end{equation}
where $\Phi(\cdot)$ is the standard normal CDF.

\paragraph{MAP + Laplace posterior.}
Let $\mathbf{f}$ be the latent utilities evaluated on the unique queried points,
with GP prior $\mathbf{f}\sim\mathcal{N}(\mathbf{0},K)$.
We estimate the posterior mode by maximizing the log posterior
\begin{equation}
\label{eq:map}
\hat{\mathbf{f}}
=
\arg\max_{\mathbf{f}}
\left[
\sum_{i=1}^{t}
\log \Phi\!\left(\frac{y_i\,\Delta_i}{\sqrt{2\lambda^2+u(c_i)^2}}\right)
-\frac{1}{2}\mathbf{f}^\top K^{-1}\mathbf{f}
\right].
\end{equation}

We then obtain an approximate posterior $(\mu_t(\cdot),\Sigma_t(\cdot,\cdot))$ with a Laplace approximation around $\hat{\mathbf{f}}$, following standard practice~\cite{rasmussen2006gpml} and UUPL~\cite{peng2025uncertainty}.

\subsection{Active Query Selection}
\label{subsec:active_query}

To reduce user effort under a fixed interaction budget, we select the next query pair
$(x^A,x^B)$ by maximizing expected information gain (IG)~\cite{mackay1992info,houlsby2011bald}.
Since the user's future confidence is unknown at selection time, we use a fixed nominal
noise scale $\sigma_{\text{nom}}$ for look-ahead evaluation.

\paragraph{Pairwise statistics under the current posterior.}
For a candidate pair, define $\delta=f(x^A)-f(x^B)$.
Under the Laplace GP posterior, $\delta$ is approximately Gaussian:
\begin{equation}
\label{eq:delta_gauss}
\delta \sim \mathcal{N}(m, v),
\end{equation}
where $m=\mu_t(x^A)-\mu_t(x^B)$ and
$v=\mathrm{Var}_t\!\big(f(x^A)-f(x^B)\big)$.

\paragraph{Information gain objective.}
Using the probit observation model with nominal noise $\sigma_{\text{nom}}$,
the predictive probability of choosing $A$ is
\begin{equation}
\label{eq:p_query}
p \triangleq P(y=+1 \mid x^A,x^B,\mathcal{D}_t)
\approx
\Phi\!\left(\frac{m}{\sqrt{v+\sigma_{\text{nom}}^2}}\right).
\end{equation}
We compute IG in a convenient 1D form (binary entropy $h(q)=-q\log q-(1-q)\log(1-q)$):
\begin{equation}
\label{eq:ig_easy}
\mathrm{IG}(x^A,x^B)
=
h\!\left(\Phi\!\left(\frac{m}{\sqrt{v+\sigma_{\text{nom}}^2}}\right)\right)
-
\mathbb{E}_{z\sim\mathcal{N}(0,1)}
\left[
h\!\left(\Phi\!\left(\frac{m+\sqrt{v}\,z}{\sigma_{\text{nom}}}\right)\right)
\right].
\end{equation}

Bıyık et al.~\cite{biyik2024active} derive a closed-form approximation for \eqref{eq:ig_easy}.

We select the next pair by
\begin{equation}
\label{eq:next_pair}
(x_{t+1}^{A},x_{t+1}^{B})
=
\arg\max_{x^A,x^B \in \mathcal{X}_{\mathrm{cand}}}
\mathrm{IG}(x^A,x^B).
\end{equation}

\subsection{Algorithmic Pipeline}
\label{subsec:pipeline}

Figure~\ref{fig:system_overview} illustrates the interaction loop, and Algorithm~\ref{alg:vpl} provides the procedural summary.
At each round $t$, the system (i) samples a candidate set $\mathcal{X}_{\text{cand}}$,
(ii) evaluates each candidate pair using the information-gain criterion (Eq.~\ref{eq:ig_easy}),
and (iii) queries the most informative pair to the user.
The user returns both a preference label $y_t$ and a confidence score $c_t$, which is mapped to the comparison noise scale via $u(\cdot)$ in the probit likelihood (Eq.~\ref{eq:probit_like}).
The GP posterior is then updated via MAP estimation (Eq.~\ref{eq:map}) with a Laplace approximation to obtain $(\mu_t,\Sigma_t)$.
After $T$ rounds, we recommend the signal $\hat{x}^*$ that maximizes the posterior mean.

\begin{algorithm}[tb]
\caption{Confidence-Aware VPL Interaction Loop}
\label{alg:vpl}
\begin{algorithmic}[1]
\Require GP prior $f\sim\mathcal{GP}(0,k)$, confidence mapping $u(\cdot)$, jitter $\lambda$, nominal noise $\sigma_{\text{nom}}$
\State Initialize $\mathcal{D}_0 \leftarrow \emptyset$
\For{$t=1$ to $T$}
  \State Sample candidate set $\mathcal{X}_{\text{cand}} \subset [0,1]^4$
  \State Select $(x_t^A,x_t^B) \leftarrow \arg\max_{x,x'\in\mathcal{X}_{\text{cand}}} \mathrm{IG}(x,x')$
  \State \textbf{User:} compare $(x_t^A,x_t^B)$, return $(y_t,c_t)$
  \State $\mathcal{D}_{t} \leftarrow \mathcal{D}_{t-1} \cup \{(x_t^A,x_t^B,y_t,c_t)\}$
  \State Update posterior via MAP (Eq.~\ref{eq:map}) and Laplace to obtain $(\mu_{t},\Sigma_{t})$

\EndFor
\State \textbf{Recommend} $\hat{x}^{*}=\arg\max_{x\in[0,1]^4}\mu_T(x)$
\end{algorithmic}

\end{algorithm}

\subsection{User Modeling}
\label{subsec:user_modeling}

To clarify the scope of our approach, we explicitly outline assumptions about user behavior, the user attributes captured by our model, and the factors currently excluded.

\paragraph{Assumptions}
We rely on two fundamental assumptions regarding the user's psychophysical state:
\begin{compactitem}
    \item \textbf{Stationarity:} We assume the user's latent preference function is approximately stationary throughout a short session (40 iterations), i.e., there is no systematic concept drift or fundamental shift in taste within this time window.
    \item \textbf{Local Smoothness:} We assume vibrotactile preferences vary smoothly with respect to the signal parameters. If a user prefers a particular setting, nearby settings in the (normalized) parameter space are expected to have similar utility, rather than exhibiting abrupt discontinuities.
\end{compactitem}

\paragraph{Modeled Aspects}
Our system explicitly captures two dimensions of the user:
\begin{compactitem}
    \item \textbf{Sensory Preference Structure:} The model learns a latent utility function $f(x)$, representing the user's subjective mapping from signal parameters to perceived utility.
    \item \textbf{Uncertainty and Calibration:} Beyond estimating \textit{what} the user prefers, we model \textit{how certain} they are. By integrating confidence scores, we treat the user as a variable-noise oracle via the mapping $u(c)$, which modulates the comparison-noise scale in the likelihood. In addition, the GP posterior provides model uncertainty over $f(x)$, supporting uncertainty-aware query selection and conservative recommendations when appropriate.
\end{compactitem}

\paragraph{Unmodeled Aspects}
The current framework focuses on within-session sensory evaluation and excludes:
\begin{compactitem}
    \item \textbf{Context Dependence:} We do not model how preferences vary with situational factors (e.g., task demands, attentional load, environment, grip style, body location) or the semantic meaning of haptic events.
    \item \textbf{Temporal Adaptation and Non-Stationarity:} The model does not account for sensory adaptation, habituation, learning effects, or systematic drift across sessions.
    \item \textbf{User Fatigue and Strategic Behavior:} We do not explicitly model fatigue-related degradation, careless responding, or strategic behavior. Behavioral signals such as response time are not utilized.
\end{compactitem}

\section{Experimental Evaluation}
\subsection{Simulation}
\label{subsec:sim}

To evaluate the algorithm's convergence without human variability, we conducted an automated simulation using a synthetic user represented by a fixed Ground-Truth (GT) utility function. 

The simulation mirrors the active learning loop used in the user study. In each iteration, the Gaussian Process (GP) selects a query pair $(x^A, x^B)$ that maximizes information gain. The simulator then acts as an oracle: it selects the option with the higher GT utility and assigns a confidence score ($1\text{--}5$) proportional to the normalized utility difference between the two options. This feedback is used to update the GP posterior.

\begin{figure}[tb]
    \centering
    \begin{subfigure}[b]{0.49\linewidth}
        \centering
        \includegraphics[width=\linewidth]{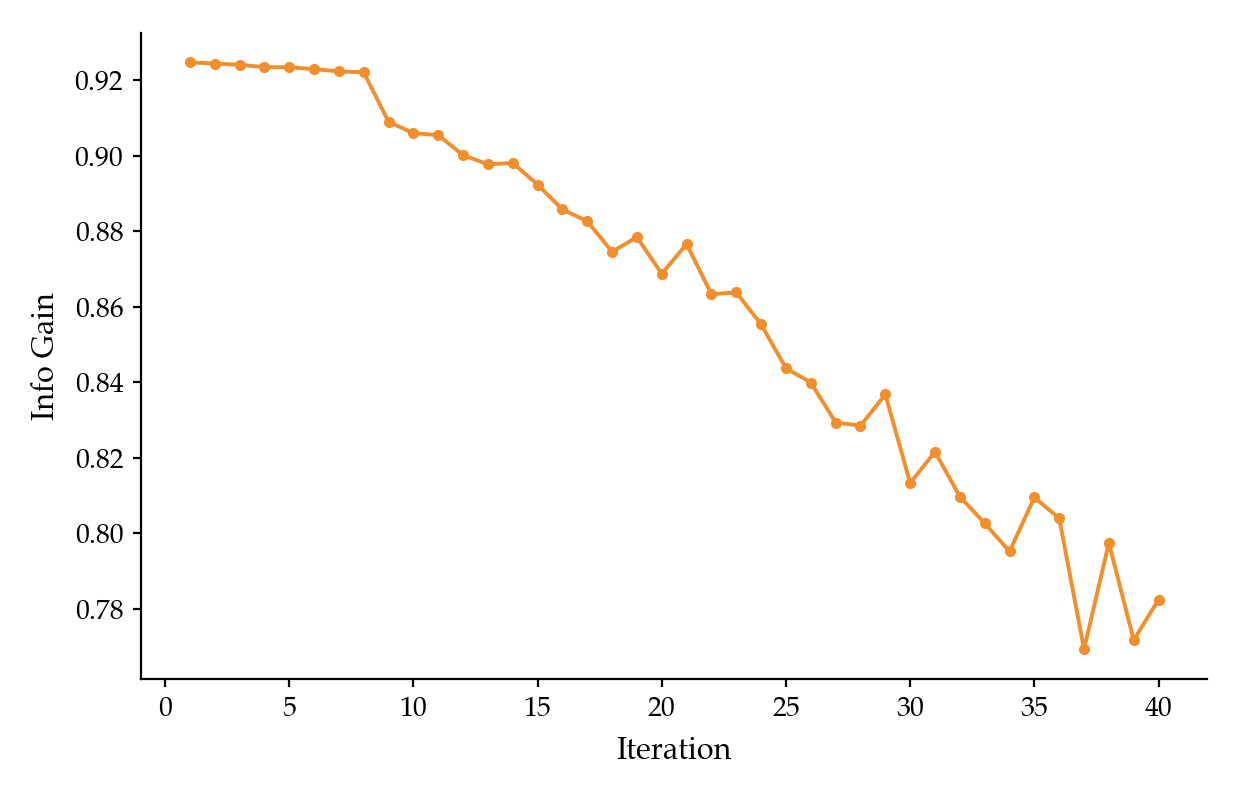}
        \vspace{-0.2in}
        \caption{Information Gain}
        \label{fig:ig_trend}
    \end{subfigure}
    \hfill
    \begin{subfigure}[b]{0.49\linewidth}
        \centering
        \includegraphics[width=\linewidth]{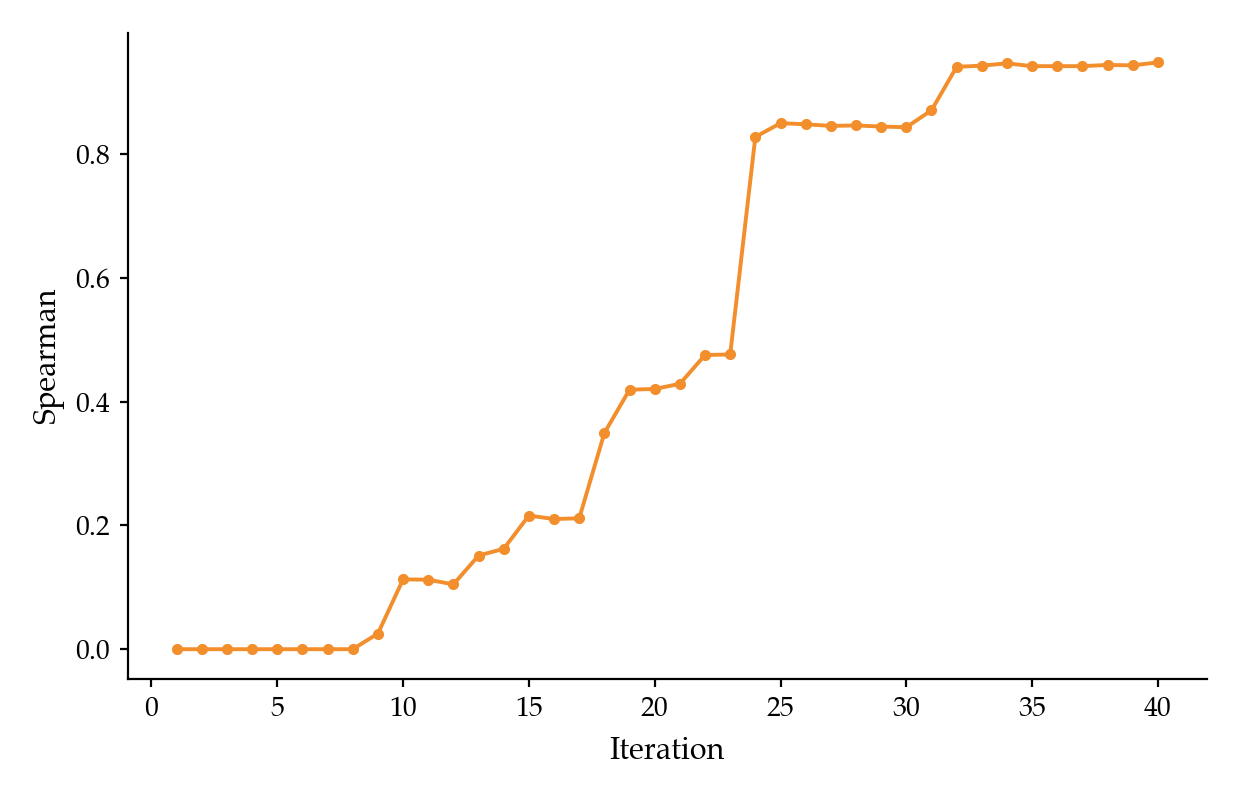}
        \vspace{-0.2in}
        \caption{Spearman Correlation}
        \label{fig:spearman_trend}
    \end{subfigure}
    \vspace{-0.15in}
     \caption{Simulation results}
    \label{fig:sim_results}
    \vspace{-1.2em}
\end{figure}

The results, shown in Fig.~\ref{fig:sim_results}, demonstrate accurate convergence. The Information Gain (Fig.~\ref{fig:sim_results}a) declines steadily, indicating that the model effectively reduces global uncertainty over time. The Spearman rank correlation between the learned model and the GT function (Fig.~\ref{fig:sim_results}b) increases and stabilizes, confirming that the algorithm accurately recovers the underlying user preferences.

\subsection{Participants}
We recruited 13 participants (P01--P13; 8 female, 5 male) aged 18 to 43 years ($M = 28.8, SD = 7.2$). 
Participants reported varying gaming frequencies (Never: 4, Occasionally: 8, Often: 1) and controller familiarity (Low: 3, Medium: 3, High: 7). The study was approved by the University Institutional Review Board, and all participants gave informed consent. 

\subsection{Experimental Procedure}
The study was conducted in a controlled environment and followed a three-step protocol: Active Learning (Phase 1), Model Validation, and Manual Adjustment (Phase 2).

\subsubsection{Phase 1: Active Preference Learning}
This phase consisted of 40 iterations aimed at training the Gaussian process (GP) model. The signal generation was initialized with a recorded time-based random seed for each session. In each iteration, participants used the left analog stick of the controller to navigate between two candidate vibrations generated as described in Section 3.1 (Option A and Option B). They could press the 'X' button to play the tactile effects as many times as needed. After comparing the signals, participants selected their preferred option and rated their decision confidence on a scale of 1 to 5. The parameters for $u$ in the human-uncertainty noise scale (Section 3.3) were calibrated via pilot studies and simulations: $\{u(1)=9.0, u(2)=3.35, u(3)=1.7, u(4)=0.66, u(5)=0.01\}$.

\subsubsection{Validation Round}
Immediately following the training phase, the system automatically generated a validation set to evaluate the quality of the learned model.

\paragraph{Stimuli Generation.}
The system first identified the signal with the highest posterior mean (denoted as $x_{\text{best}}$). It then sampled 6 additional points from the parameter space, enforcing a minimum distance constraint (threshold = $0.05 \times \text{diagonal length}$) to ensure diversity. The posterior means of these 7 points (1 best + 6 sampled) were calculated to facilitate specific pairwise comparisons.

\paragraph{Comparison Tasks.}

To evaluate the accuracy of the learned preference model, participants were asked to perform a set of validation tasks. The system generated 4 specific comparison pairs based on the learned utility function. For each pair, participants compared the two signals (labeled Option A and Option B) and selected their preferred one. The goal was to determine if the participants' actual choices aligned with the model's predicted rankings. The pairs were designed to test different aspects of the model:

\begin{compactenum}
    \item \textbf{Anchor (Easy):} $x_{\text{best}}$ vs. $x_{\text{worst}}$. 
    Sanity check to confirm the participant prefers the optimal signal over the worst one.

    \item \textbf{Anchor (Medium):} $x_{\text{best}}$ vs. $x_{\text{mid}}$. 
    Tests whether the model can correctly distinguish the optimal signal from a medium-utility alternative.

    \item \textbf{Global Trade-off:} Two signals with a large Euclidean distance and significant utility difference. 
    Evaluates if the model correctly predicts preferences between two physically very different signals.

    \item \textbf{Consistency Check:} A repetition of the ``Anchor (Medium)'' pair with the positions of A and B swapped. 
    Detects potential position bias or random guessing by the participant.
\end{compactenum}



\subsubsection{Phase 2: Free-Adjustment Task}
To obtain a set of user-defined optima as a high-quality reference, participants entered a free-adjustment phase. They were presented with a graphical interface allowing free control over the four vibration parameters (intensity, motor balance, rhythm, grain). Participants were asked to explore the space and manually tune the parameters to create their ideal vibration. This process was repeated three times, and the top three favorite configurations were recorded.

\subsection{Measures}
We employed standard psychometric scales, custom questionnaires, and system logs to evaluate task workload and system performance.

\paragraph{System Logs and Objective Data.}
Throughout the experiment, the system automatically logged detailed interaction data for quantitative analysis:
\begin{compactitem}
    \item \textbf{Interaction History:} For all 40 training iterations and 4 validation rounds, we recorded the physical parameters of the candidate pairs $(x^A, x^B)$, the participant's binary choice ($y \in \{A, B\}$), the self-reported confidence level, and the calculated Information Gain (IG) for each query.
    \item \textbf{Convergence Data:} We logged the model's final recommended signal ($x_{\text{rec}}$) derived from the posterior mean after the active learning phase.
    \item \textbf{User-Defined Favorites:} We recorded the parameters of the top 3 "favorite" signals manually created by the participants in Phase 2, along with their creation timestamps. These serve as the reference standard for evaluating the model's recommendation accuracy.
\end{compactitem}

\paragraph{Subjective Workload (NASA-TLX).}
Participants completed the NASA Task Load Index (NASA-TLX) to assess the workload across six dimensions~\cite{hart1988nasatlx}: Mental Demand, Physical Demand, Temporal Demand, Performance, Effort, and Frustration. Each dimension was rated on a scale from 0 (Low) to 20 (High).

\paragraph{User Experience and System Usability.}
We designed a custom 7-point Likert scale questionnaire (1 = Strongly Disagree, 7 = Strongly Agree) to evaluate six specific aspects of the system:
\begin{compactitem}
    \item \textbf{Ease of Use:} "The system was easy to use"
    \item \textbf{Controllability:} "I understood what each parameter controls"
    \item \textbf{Decision Support:} "The uncertainty rating helped me make decisions"
    \item \textbf{Satisfaction:} "How satisfied are you with the final recommended signal?"
    \item \textbf{Expectation Match:} "Did the recommended signal match your expected preferred haptic feeling?"
    \item \textbf{Comfort:} "Overall comfort was acceptable"
\end{compactitem}

\section{Results}
\label{sec:results}

In this section, we present the evaluation results to answer our three research questions: (1) Can the system learn a accurate, individualized haptic preference distribution within a limited number of interaction rounds? (\textbf{RQ1}); (2) Are users satisfied with the signals recommended by the system? (\textbf{RQ2}); and (3) Are the interaction workload and usability within an acceptable range? (\textbf{RQ3}). 


\subsection{Objective Metrics (RQ1)}
\label{sec:results_objective}

\subsubsection{Held-out Validation Accuracy}
To assess whether the learned model generalizes to unseen data, we analyzed the prediction accuracy on the held-out validation set. 

\begin{figure}[t]
  \centering
  \includegraphics[width=\linewidth]{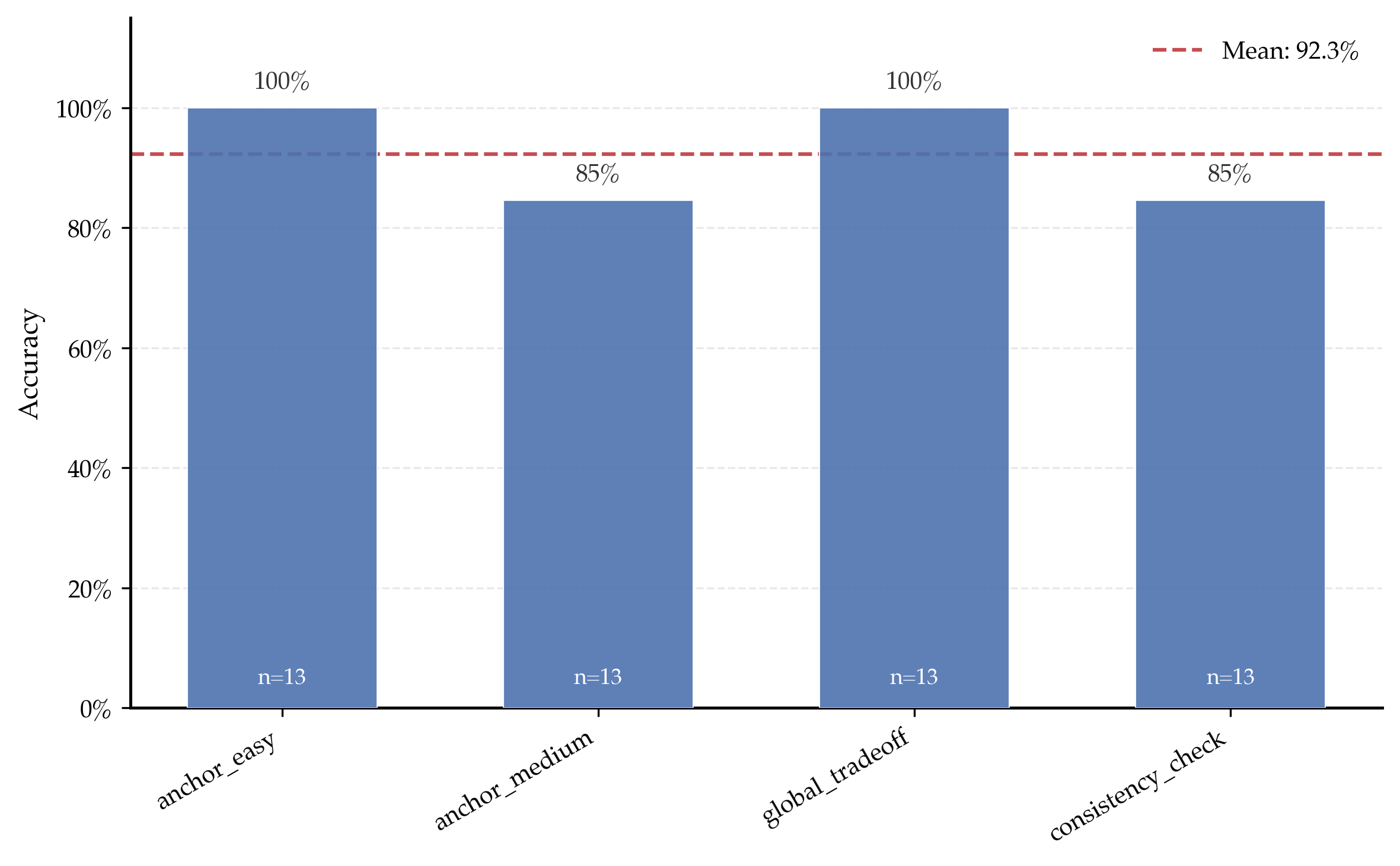}
  \vspace{-0.3in}
  \caption{Validation accuracy grouped by item type}
  \label{fig:val_by_item}
  \vspace{-1em}
\end{figure}

\begin{figure}[t]
  \centering
  \includegraphics[width=\linewidth]{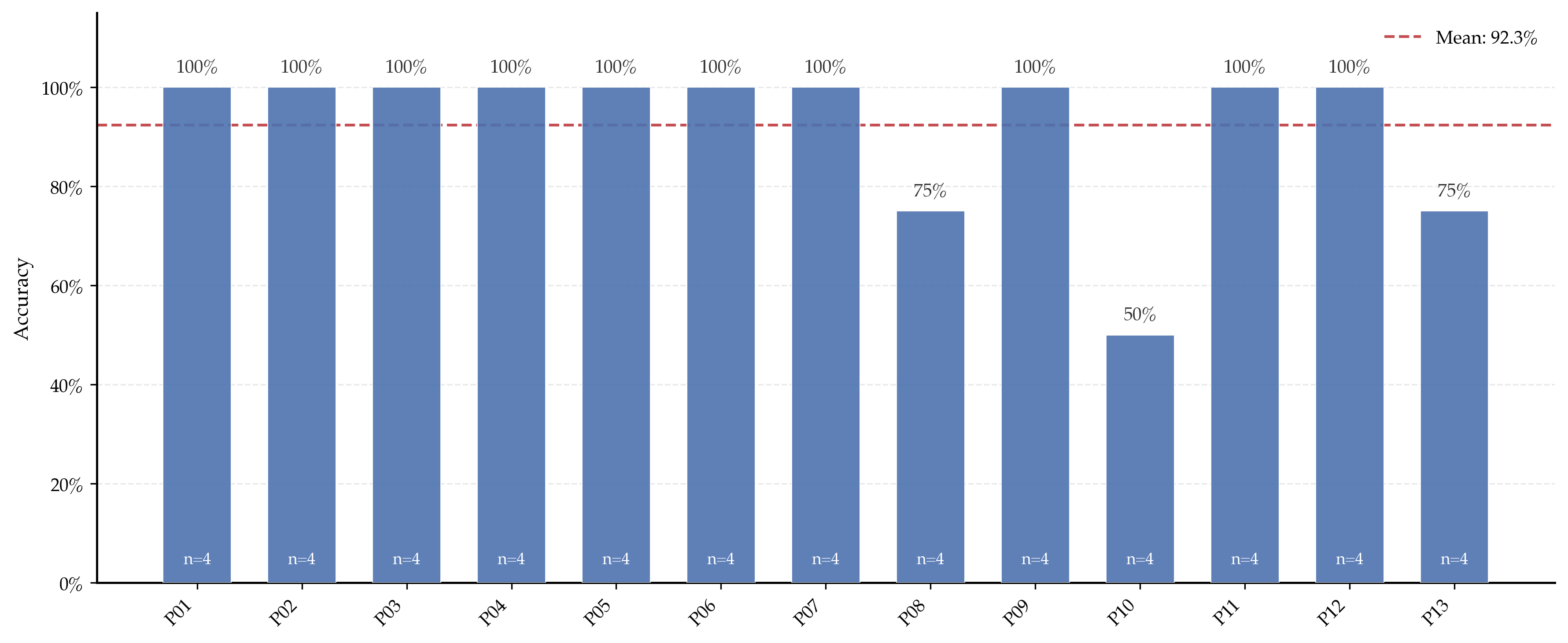}
  \vspace{-0.2in}
  \caption{Validation accuracy grouped by participant}
  \label{fig:val_by_part}
  \vspace{-1em}
\end{figure}

Figure~\ref{fig:val_by_item} summarizes the validation accuracy grouped by item type, and Figure~\ref{fig:val_by_part} shows the results grouped by participant. The overall mean pairwise accuracy reached \textbf{92.3\%} (range: 85\%--100\%) with a mean user confidence of \textbf{4.38/5}. This high accuracy demonstrates that the GP model successfully captured the latent structure of participants' haptic preferences and could reliably predict choices in held-out comparisons.

\subsubsection{Consistency with User-Defined Optima}
\label{sec:results_best_signal_consistency}
To further verify the model's validity, we examined the alignment between the model's predictions and the user-defined preferences. We took the ``best'' signal manually tuned by each participant in Phase 2 and evaluated its standing within their learned preference model. Specifically, we calculated the percentile rank of this user-defined signal within the model's posterior mean distribution over the entire search space.
The results show that the user-defined best signals achieved an average percentile rank of \textbf{81.2\%}. This indicates a strong alignment: the signals that users subjectively felt were ``best'' were also predicted by the model to be in the top tier (top $\approx$19\%) of the utility space.

\vspace{0.5em}
\noindent\textbf{Summary for RQ1:} Supported by the high held-out validation accuracy (92.3\%) and the strong consistency with user-defined optima (81.2\% percentile rank), we conclude that the system is capable of learning a stable and accurate individual haptic preference distribution within 40 iterations.

\subsection{Subjective Evaluation (RQ2 \& RQ3)}
\label{sec:results_subjective}

\subsubsection{System Satisfaction (RQ2)}
\begin{figure}[tb]
  \centering
  \includegraphics[width=0.8\linewidth]{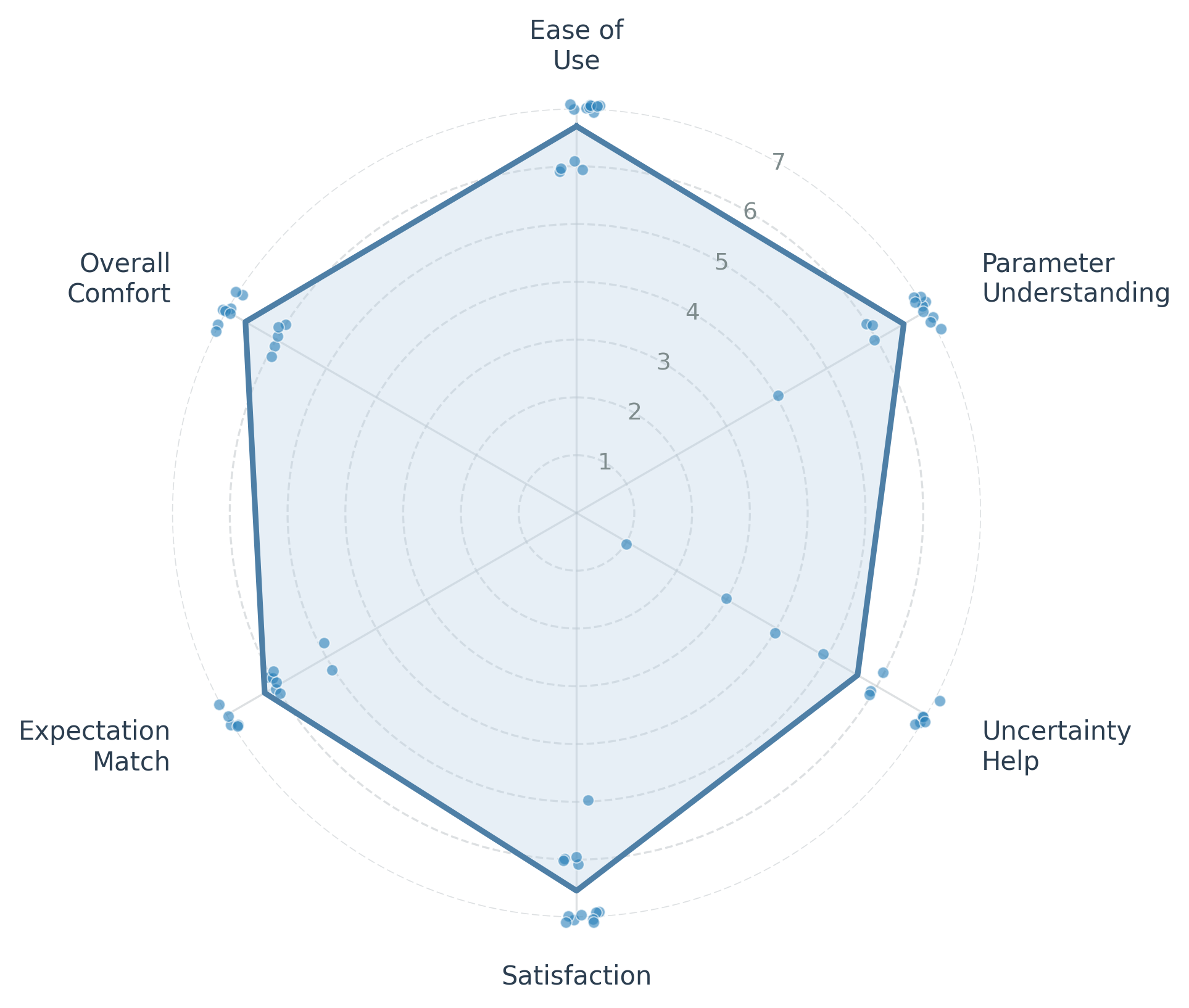}
  \vspace{-0.15in}
  \caption{Post-study subjective ratings (N=13)}
  \label{fig:ev}
  \vspace{-1.5em}
\end{figure}

We evaluated user satisfaction through a post-study questionnaire using a 7-point Likert scale (1=Strongly Disagree, 7=Strongly Agree). As shown in Figure~\ref{fig:ev}, participants reported high ratings for \textbf{Usability} ($M=6.69, SD=0.48$) and \textbf{Comfort} ($M=6.62, SD=0.51$). \textbf{Controllability} ($M=6.54, SD=0.88$) and \textbf{Overall Satisfaction} ($M=6.54, SD=0.66$) were also highly rated.
Notably, the \textbf{Expectation Match} score ($M=6.23, SD=0.73$) suggests that the algorithmically recommended signals closely matched users' mental models of their preferred haptics. The \textbf{Decision Support} rating showed higher variability ($M=5.62, SD=1.89$). This is because the uncertainty help operates on the final recommendation generation, without being explicitly perceived by users as an improvement during the interaction.

\subsubsection{Workload Assessment (RQ3)}
\begin{figure}[tb]
  \centering
  \includegraphics[width=\linewidth]{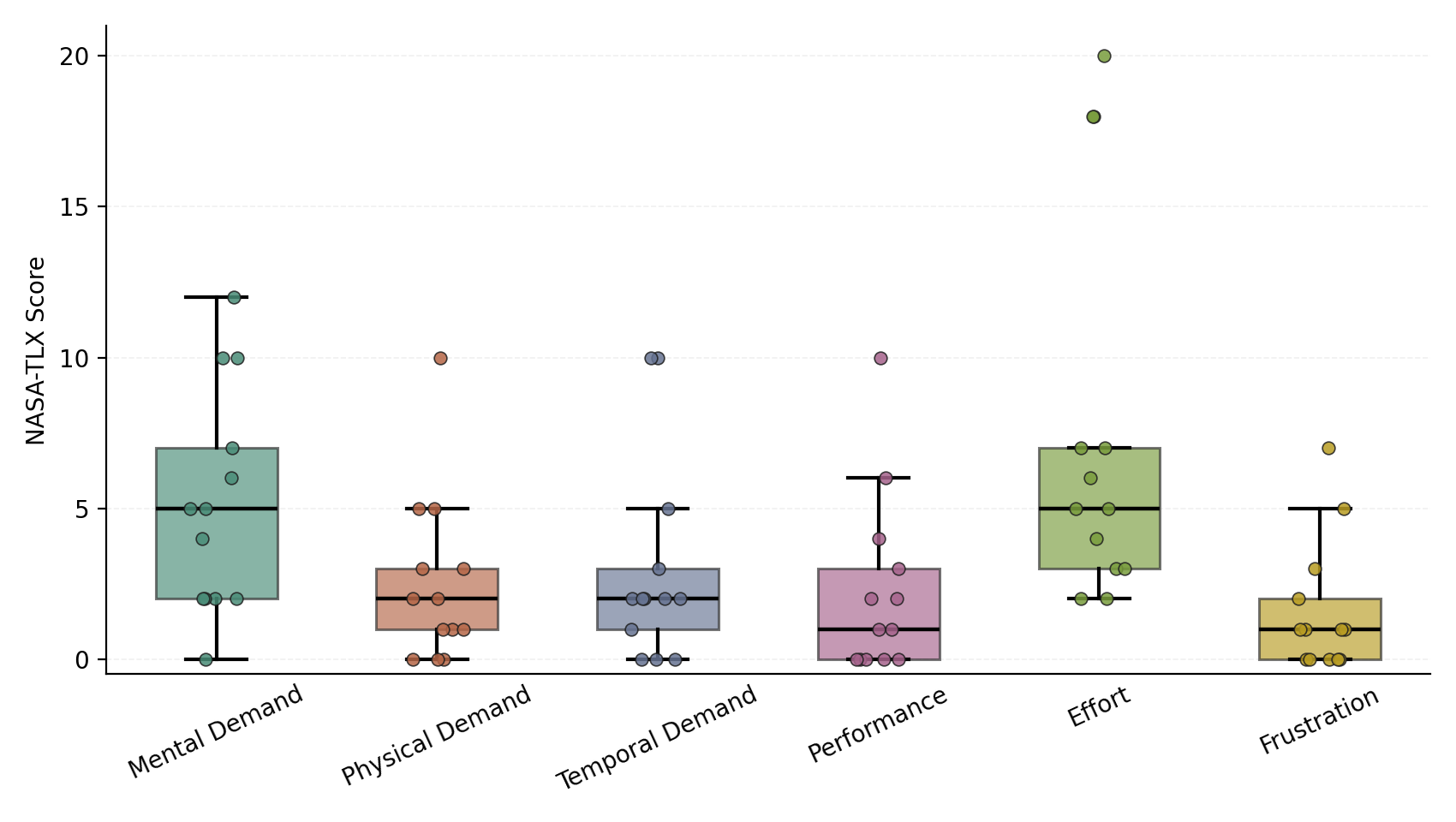}
  \vspace{-0.3in}
  \caption{NASA-TLX workload ratings.}
  \label{fig:nasa}
  \vspace{-1.7em}
\end{figure}

The active-learning phase lasted \textbf{763.31 $\pm$ 188.79 s} (\textbf{12m43s $\pm$ 3m09s}), including repeated playback, showing that the 40-comparison budget remained a short, bounded session.
We assessed interaction workload using the NASA-TLX scale (0--20). Figure~\ref{fig:nasa} presents the scores across six dimensions. The overall workload was low ($M=3.71 \pm 1.80$).
Detailed analysis reveals that \textbf{Effort} ($M=7.69 \pm 6.49$) and \textbf{Mental Demand} ($M=5.15 \pm 3.72$) were the highest among dimensions, reflecting the cognitive attention required to distinguish subtle haptic differences. Conversely, \textbf{Physical Demand} ($M=2.54 \pm 2.82$), \textbf{Temporal Demand} ($M=3.00 \pm 3.39$), and \textbf{Frustration} ($M=1.62 \pm 2.18$) remained very low. 
Crucially, the \textbf{Performance} score was low ($M=2.23 \pm 2.98$). Note that on the NASA-TLX scale, a lower Performance score indicates higher self-perceived success (i.e., less failure). This confirms that participants felt successful in guiding the system to their preferred outcomes.

\vspace{0.5em}
\noindent\textbf{Summary for RQ2 \& RQ3:} The high satisfaction scores confirm that users were satisfied with the recommended signals (RQ2). The NASA-TLX results demonstrate that despite the inherent mental effort required for psychophysical comparisons, the overall workload and usability remained within an acceptable range (RQ3).

\section{Discussion}
\subsection{Sampling Strategy in Haptics}
\label{sec:discussion_eubo_infogain}

A key design choice in preference-based interactive haptics is the \emph{sampling strategy}: which pair of signals to query next under a limited interaction budget. In this work, we adopt an \textbf{information-gain (InfoGain)} strategy that is \emph{model-driven}: it prioritizes queries expected to maximally reduce uncertainty in the learned preference model, leading to broader exploration especially in regions where the model is uncertain or user judgments are inconsistent.

In contrast, \textbf{Expected Utility of the Best Option (EUBO)} is a commonly used acquisition function in preferential Bayesian optimization that is \emph{solution-driven}~\cite{lin2022bope,astudillo2023qeubo}. Rather than optimizing overall model fidelity, EUBO prioritizes query pairs that are expected to most improve the quality of the \emph{final recommended signal} (i.e., the best candidate under the current posterior). Consequently, EUBO tends to more aggressively allocate samples densely around the currently promising region to refine a local neighborhood and reach a high-utility optimum quickly.

\begin{figure}[tb]
  \centering
  \begin{subfigure}[t]{0.49\linewidth}
    \centering
    \includegraphics[width=\linewidth]{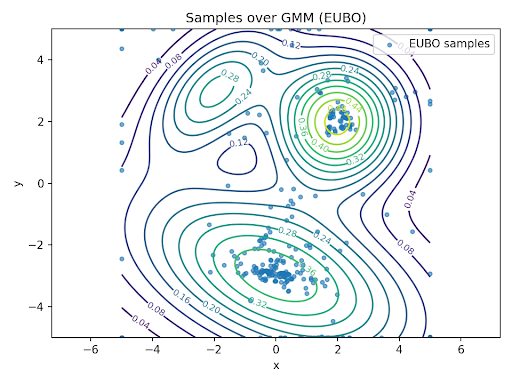}
    \vspace{-0.2in}
    \caption{EUBO sampling}
    \label{fig:fig8_eubo}
  \end{subfigure}
  \hfill
  \begin{subfigure}[t]{0.49\linewidth}
    \centering
    \includegraphics[width=\linewidth]{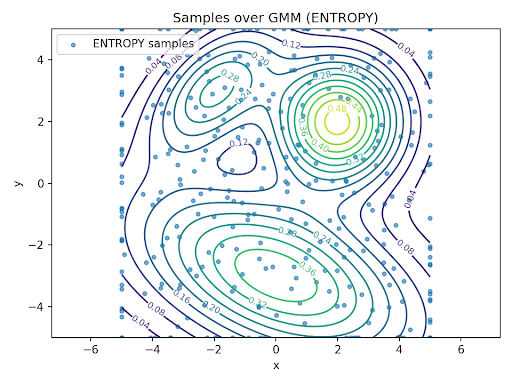}
    \vspace{-0.2in}
    \caption{InfoGain-style sampling}
    \label{fig:fig9_entropy}
  \end{subfigure}
  \vspace{-0.15in}
  \caption{Comparison of sampling behaviors on GMM} 
  \label{fig:sampling_comparison}
  \vspace{-1em}
\end{figure}

\begin{figure*}[t]
  \centering
  \includegraphics[width=\textwidth, trim=0cm 0.5cm 0cm 0.5cm, clip]{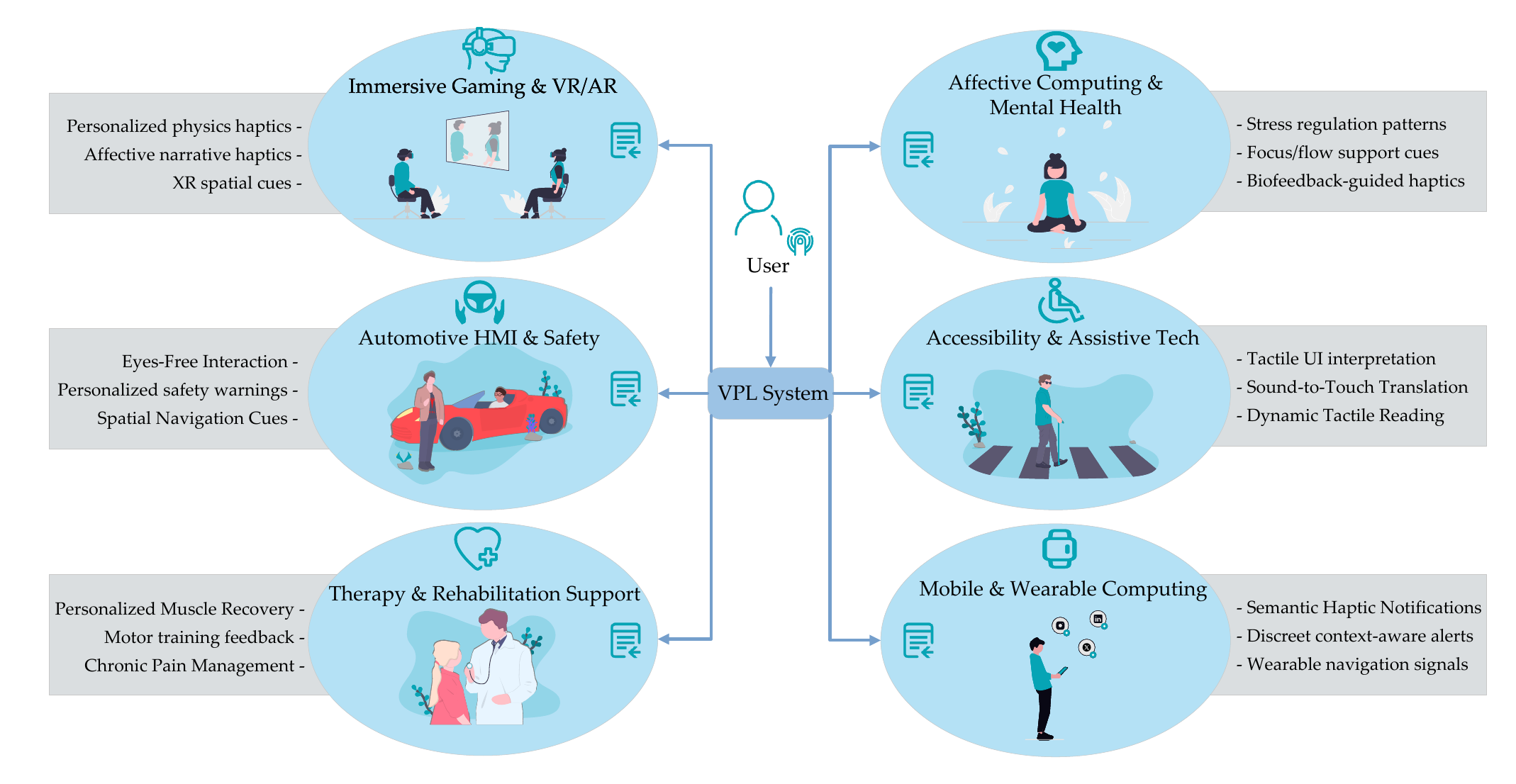} 
  \vspace{-0.3in}
  \caption{Illustrative application domains where VPL could support personalized vibrotactile feedback.}
  \label{fig:applications}
\vspace{-1em}
\end{figure*}

Figure~\ref{fig:sampling_comparison} contrasts sampling behaviors on a multimodal landscape: EUBO concentrates near a single high-utility mode, while InfoGain explores broadly to resolve global structure. This leads to different trade-offs for haptic applications. For \emph{short, goal-oriented sessions} (e.g., quickly tuning a single ``best-feeling'' vibration style for a game event), EUBO can quickly converge to a satisfying signal. However, due to perceptual indistinguishability and hardware-limited resolution, it may waste budget on low-information comparisons near an apparent optimum, increasing user fatigue. In contrast, InfoGain better supports \emph{robust and reliable} personalization by learning a calibrated preference landscape and identifying uncertain or indistinguishable regions, enabling stable recommendations and cross-context reuse (multiple haptic events) at the cost of slower convergence to a single optimum under a fixed budget.

Overall, EUBO and InfoGain represent complementary priorities: \emph{fast best-signal optimization} vs.\ \emph{faithful preference learning under uncertainty}. A promising direction is a hybrid schedule: use InfoGain early to map the preference structure and detect indistinguishable regions, then switch to EUBO refinement to polish the final recommendation.

\subsection{Applications}
\label{subsec:application}

Fig.~\ref{fig:applications} outlines domains where VPL can support personalized vibrotactile feedback. We emphasize two deployment-relevant advantages: \textbf{(i) robustness under hard constraints} and \textbf{(ii) uncertainty awareness that avoids ambiguous regions}.

\paragraph{Advantage 1: Robustness under hard constraints.}
Instead of optimizing only a single local optimum, VPL learns an uncertainty-calibrated \emph{global} preference landscape. When deployments impose hard limits (e.g., maximum intensity, restricted frequency ranges, bandwidth limits, comfort/safety caps, or distinctiveness requirements across alerts), VPL can \emph{directly} choose a strong recommendation \textbf{within the feasible set}, without re-running optimization or preference elicitation.

\textbf{Automotive HMI \& safety.}
Vehicle haptics are safety-critical and constrained by intensity caps, timing limits, and \emph{separability} across warning types (e.g., lane departure, collision, navigation), which may vary by context. When certain designs are infeasible, VPL can identify a strong feasible alternative from its global model.

\textbf{Mobile \& wearables.}
Haptics often encode multiple alerts under intensity/duration caps, limited actuator bandwidth/OS scheduling, and explicit \emph{distinctiveness} requirements. The goal is a usable set of preferred, discriminable signals; VPL can propose alternatives when a preferred design violates constraints or conflicts with other alerts, maintaining quality under changing limits (e.g., silent mode, workouts, long-wear comfort).

\textbf{Therapy \& rehabilitation.}
Wellbeing applications impose conservative safety bounds and user tolerance may change over time. VPL supports constraint-aware personalization within a safe envelope and provides stable fallbacks when constraints tighten, enabling graded programs instead of over-focusing near a potentially unreliable local optimum.

\textbf{Immersive gaming and VR/AR.}
These domains benefit from fast personalization of event-specific haptic styles, especially when multiple events must be pleasant yet mutually distinguishable.

\paragraph{Advantage 2: Uncertainty awareness to avoid ambiguous regions.}
VPL explicitly models uncertainty in user judgments, which is common due to perceptual indistinguishability and context effects. It learns not only where utility is high but also where confidence is high, allowing it to avoid low-confidence regions. This is useful in affective computing and mental health support, where a conservative, lower-risk pattern is preferable when judgments are ambiguous or preferences fluctuate.

\subsection{Limitations}
\label{subsec:limitations}

\paragraph{Limited consistency under a finite query budget.}
With a fixed budget of 40 interactions, the learned preference landscape should be interpreted as a \emph{low-resolution} approximation. Very similar vibrotactile signals may be perceptually indistinguishable, and comparisons between nearby points in the parameter space are dominated by response noise, limiting fine-grained consistency.

\paragraph{Restricted parameterization and scalability.}
We model vibrations using four global parameters and a fixed 3 s duration, which supports efficient interaction but excludes richer spatiotemporal patterns (e.g., complex envelopes or multi-actuator effects). Extending VPL to higher-dimensional or compositional signal spaces is promising but would increase both computational cost and required user interactions.

\paragraph{Subjective confidence and fixed noise mapping.}
Our method depends on self-reported confidence and a fixed confidence-to-noise mapping. Because confidence is subjective and may vary across users or over time, a single mapping may be suboptimal. Future work could adapt or personalize this mapping through calibration or online learning.

\section{Conclusion}


We presented Vibrotactile Preference Learning (VPL), an interactive method that personalizes vibrotactile feedback using simple A/B comparisons, together with users' self-reported confidence. VPL uses the confidence ratings to decide how much to trust each response, and it chooses the next comparison to ask so that each interaction is as informative as possible. In a user study ($N=13$) with an Xbox controller, VPL reached \textbf{92.3\%} accuracy on the validation set after \textbf{40} rounds, with an average confidence of \textbf{4.38/5}, and participants reported high satisfaction and low workload. Future work will extend VPL to richer spatiotemporal haptic patterns and make the approach easier to scale and deploy in real-world systems.

\clearpage 
\bibliographystyle{ACM-Reference-Format}
\bibliography{references}

\appendix

\end{document}